\documentstyle[aps,prd,preprint,tighten,floats,epsf]{revtex}

\begin{document}

\title{Construction of a robust warm inflation mechanism}

\author{
Arjun Berera$^{1}$\thanks{E-mail address:
ab@ph.ed.ac.uk} and
Rudnei O. Ramos$^{2}$\thanks{E-mail address: rudnei@uerj.br}}

\address{
{\it $^{1}$School of Physics,
University of Edinburgh, Edinburgh EH9 3JZ, Great Britain} \\
{\it  $^{2}$Departamento de F\'{\i}sica Te\'orica,
Universidade do Estado do Rio de Janeiro,
20550-013 Rio de Janeiro, RJ, Brazil}}

\maketitle

\begin{abstract}

A dissipative mechanism is presented, which emerges
in generic interacting quantum field systems and which
leads to robust warm inflation. An explicit example is
considered, where using typical 
parameter values, it is shown that
considerable radiation can be produced during
inflation. The extension of our results to expanding
spacetime also is discussed.

\medskip

PACS numbers:
98.80.Cq, 05.70.Ln, 11.10.Wx
\end{abstract}

\medskip

\noindent
keywords: cosmology, warm inflation, dissipation, field dynamics

\bigskip

In Press Physics Letters B, 2003

\medskip \medskip

\section{Introduction}
\label{intro}

Inflationary dynamics inherently is a multifield problem, since the
vacuum energy that drives inflation eventually must convert to
radiation, which generally is comprised of a variety of particle
species. Phenomenologically it has been shown that the inflation and
radiation production phases can be two well separated periods in
scenarios generically termed supercooled (or isentropic)
inflation (for a review see
\cite{olive}), or radiation production can occur concurrently with
inflationary expansion in scenarios generically termed warm 
(or nonisentropic) inflation
\cite{wi}. Warm inflation is a broader picture, since the extent of
radiation production during inflation is variable, so that supercooled
inflation emerges as the limiting case of zero radiation production. 

Although by now considerable work has demonstrated its phenomenological
significance \cite{phenomen}, one key barrier to the warm inflation
picture has been establishing plausibility of its dynamics from first
principles quantum field theory. To some extent this point has been
overemphasized for warm inflation, since in similar respects particle
production during the far out-of-equilibrium reheating phase of
supercooled inflation is not well understood, thus leaving
incompleteness also to this picture. However for supercooled inflation,
since particle production is assumed not to affect large scale structure
formation during inflation, thus the main observational predictions,
these shortcomings are cast aside as secondary concerns. Nevertheless,
without a solution here, this picture is unproven. On the other hand,
the warm inflation picture makes no {\it a priory} assumption that
particle production does not affect large scale structure formation. As
such, the particle production problem appears more acute here. More
basically a proper understanding of particle production should mean that
theory itself can decide which or to what extent either of these two
pictures is valid. Undoubtedly, no theory based on inflationary
expansion will ever emerge, until particle production in quantum field
theory is adequately understood.

This is a major problem, which must be tackled in steps. {}Fair enough
is to attempt to see how well either picture of inflation can be
understood from first principles and {\it en route} hope a clearer
general picture eventually will emerge. {}For warm inflation, there is
greater possibility to understand particle production, and eventually
reach closure at a theoretical level about the viability of this picture
as a description of the early universe. The reason is that recall in
this picture the scalar inflaton field is required to have a slow,
overdamped motion. As such, adiabatic methods of quantum field theory
are applicable here, and these are the only methods for which
dissipation can be unarguably analyzed.

The road toward a first principles warm inflation picture primarily has
been hindered by basic gaps in the understanding of dissipative quantum
field theory, which during the course of developing warm inflation are
being filled \cite{bgr1,bgr2,yl,br,ian2,HR}. 
The first attempt to understand warm
inflation dynamics utilized finite temperature dissipative quantum field
theory, since some formalism already existed here
\cite{gr,morikawa,ms1,hs,ring1}. Based on this work \cite{bgr1},
statements of a general sort have been made about the impossibility of
warm inflation dynamics \cite{yl}. However, these criticisms failed to
recognize that the key problems were specific to the restrictive
constraints of the high-$T$ approximation and were not reflexive of warm
inflation in general.

Intrinsically, warm inflation is an out-of-equilibrium problem, in that
it is not tied to any specific equilibrium statistical state, but rather
simply requires radiation production concurrent with the overdamped
relaxation of a global order parameter. Although the actual statistical
state during warm inflation may not be very far from an equilibrium
state, at present the problem is simply technical limitations
in describing the scope of such states.
{}Furthermore, as has been noted
\cite{wi,br}, very little radiation production
during inflation, at the scale of tens of
orders of magnitude below the vacuum energy density, is
already sufficient to affect large scale structure formation and create
an adequately high post-inflation temperature.

With these thoughts in mind, in \cite{br} a simple attempt was made to
circumvent the specific constraints of the high-temperature formalism,
by examining dissipation at zero-temperature. The point there was to
investigate a suggestions learned from our high-temperature analysis,
that alleviation of the constraints specific to the high-$T$
approximation would adequately allow realizing robust radiation
production during warm inflation.
The main purpose of \cite{br} was to develop the necessary formalism, but in
addition one suggestive mechanism was identified that could realize this
point, which involved a scalar $\Phi$ field (whose zero mode can be associated,
e.g., with the inflaton) exciting heavy
$\chi$-bosons which then decay into lighter $\psi$-fermions. This letter
reports a detailed investigation of this process and
demonstrates that it is a robust mechanism for warm inflation. {}For
this, in Sec. \ref{sect2} a linear response 
derivation will be presented, which in the
adiabatic regime and at leading order is equivalent to the closed time
Lagrangian formalism, but is simpler and physically more transparent.
Then in Sec. \ref{sect3} an alternative derivation is presented,
using canonical methods. 
{}From this approach, the origin of 
particle production and energy balance for this
mechanism will be clarified. 
Next, Sec. \ref{sect4} gives a physical picture
to the mechanism and supplies an explicit numerical example to demonstrate
the extent of radiation production it yields during inflation. 
Sec. \ref{sect5} discusses the extension of the calculation
to expanding spacetime.  {}Finally the conclusions are given
in Sec. \ref{concl}.

\section{A model for robust radiation production}
\label{sect2}

We consider a multi-field model, first studied in \cite{br}, of a
scalar field $\Phi$ interacting with a set of scalar fields $\chi_j$,
$j=1, \ldots, N_\chi$, which in turn interact with fermion fields
$\psi_k$, $k=1, \ldots, N_\psi$, with Lagrangian density

\begin{eqnarray}
\lefteqn{{\cal L} \! = \!\! \frac{1}{2} (\partial_\mu\Phi)^2 \! - \!
\frac{m_{0\phi}^2}{2}\Phi^2 \! -\frac{\lambda}{4 !} \Phi^4
+ \! \sum_{j=1}^{N_{\chi}} \left[ \frac{1}{2} (\partial_\mu\chi_j)^2 \! - \!
\frac{m_{0\chi_j}^2}{2}\chi_j^2 \right.} \nonumber \\
& &-  \!\left. \frac{f_j}{4 !} \chi_j^4 \!-\! \frac{g_{j}^2}{2}
\Phi^2 \chi_{j}^2 \!
\right]\! +\! \sum_{k=1}^{N_{\psi}}
\bar{\psi}_{k} \! \left[i \not\!\partial \!-\! m_{0\psi_k}
\!-\! \sum_{j=1}^{N_\chi} h_{kj} \chi_j \! \right] \psi_{k} \;.
\label{model}
\end{eqnarray}
The regime of interest for warm inflation, that is studied here is
$m_{\chi_j} > 2 m_{\psi_k} > m_{\phi}$, where these are
the renormalized and, if relevant, background field dependent
masses.

By decomposing $\Phi$ in terms of a 
homogeneous classical part, $\varphi(t)$, and its fluctuations 
$\phi$, the effective equation of motion (EOM)
for $\varphi$ emerges as

\begin{eqnarray}
&&\ddot{\varphi}(t) + m_{0\phi}^2 \varphi(t) + \frac{\lambda}{6} \varphi^3(t)
+\frac{\lambda}{2} \varphi(t) \langle \phi^2 \rangle
+\frac{\lambda}{6} \langle \phi^3 \rangle \nonumber \\
&&+ \sum_{j=1}^{N_{\chi}} g_j^2 \left[\varphi (t) \langle \chi_j^2 \rangle +
\langle \phi \chi_j^2 \rangle \right] =0 \;.
\label{eqphi}
\end{eqnarray}
We will use a linear response theory approach in which the field
averages in Eq. (\ref{eqphi}) are expressed in terms of the respective
field propagators $G_\phi(x,x')$ and $G_{\chi_j} (x,x')$.
Also in the following, we derive the $\varphi$ effective EOM from an adiabatic 
approximation.
This approximation requires
that all macroscopic motion is slow relative to the characteristic
scales of the microscopic dynamics. In our model the time scale for
microscopic dynamics is represented through the (inverse of the)
particle decay widths
$\Gamma_{\phi}$, $\Gamma_{\chi}$ and for macroscopic dynamics is
contained in $\varphi(t)$, with the basic consistency condition
\cite{bgr1}

\begin{equation}
\dot \varphi/\varphi \ll \Gamma_{\phi}, \Gamma_{\chi} \;.
\label{adapp}
\end{equation}

Turning to the derivation,  consider first
$\langle \chi^2_j \rangle$. This expectation value
can be expressed in terms of the coincidence
limit of the (causal) two-point
Green's function for the $\chi_j$ field, $G^{++}_{\chi_j} (x,x')$.
Recall that this Green's function is the $(1,1)$-component
of the real time matrix of full
propagators, all of which satisfy the appropriate Schwinger-Dyson equations
(see, {\it e.g.}, \cite{bgr1,br} for additional details)
\begin{eqnarray}
\lefteqn{\left[\Box + m_{\chi_j}^2 + g_j^2 \varphi^2(t) \right]
G_{\chi_j} (x,x')} \nonumber \\
 && ~~~~~~~~~~~~+ \int d^4 z \Sigma_{\chi_j} (x,z) G_{\chi_j} (z,x') =i \delta(x,x')  \;,
\label{Gchi}
\end{eqnarray}
\noindent
where $\Sigma_{\chi_j}$ is the $\chi_j$ field self-energy.
The field frequencies appearing in these propagators
depend on the background field configuration $\varphi(t)$. This
field is decomposed as $\varphi(t)
= \varphi_0 + \delta \varphi(t)$, where $\varphi_0$ is a constant (the
value of the field at say the initial time $t=t_0$) and $\delta \varphi(t)$
is treated perturbatively. This is just a
linear response theory approach to calculating
the averages of the fields appearing in Eq. (\ref{eqphi}).
{}Following this procedure,  we have that $\langle \chi^2_j \rangle$
can be written to lowest order as
\begin{eqnarray} 
\lefteqn{\langle\chi_j^2 \rangle \simeq \langle \chi_j^2 \rangle_0 }\nonumber \\
&& ~~~ -
i \int_{-\infty}^t 
dt' \frac{g_j^2}{2} \left[\varphi^2 (t') - \varphi_0^2 \right] 
\langle [\chi_j^2({\bf x},t),\chi_j^2({\bf x},t')]
\rangle\;, 
\label{response}
\end{eqnarray}
where $\langle \ldots \rangle_0$ means the correlation function
evaluated at the initial time. 
The $\varphi^2$ dependence in Eq. (\ref{response})
emerges from expanding the two point function
with respect to the $\delta \varphi$
dependent terms.  {}Formally this can
be done by treating $\delta \varphi$ dependent
terms in the shifted potential as interaction vertices.
This implies adding an interacting vertex quadratic in the $\chi_j$ field,
with Feynman rule $-i g_j^2/2 \; [\varphi^2(t) - \varphi_0^2]$,
and is used in calculating
the leading order one-loop bubble diagram that gives the two-point
function.
This method was first implemented to study dissipation in
\cite{ms1,hs} and more recently
in \cite{br}. This is also analogous to the functional
Schwinger closed time path formalism used in \cite{gr,bgr1}.
Using translational invariance we can now write
$\langle [\chi_j^2({\bf x},t),\chi_j^2({\bf x},t')]
\rangle$, appearing in Eq. (\ref{response}), in terms of the  
two-point
Green's function for the $\chi_j$ field, $G^{++}_{\chi_j} (x,x')$, as

\begin{eqnarray}
\lefteqn{\langle [\chi_j^2({\bf x},t),\chi_j^2({\bf x},t')]
\rangle = 2 i \: {\rm Im} \langle T \chi_j^2({\bf x},t)\chi_j^2({\bf x},t')\rangle}
\nonumber \\
&&~~~~~~~~~~~~= 4 i \:\int \frac{d^3q}{(2 \pi)^3}
{\rm Im}[G_{\chi_j}^{++}({\bf q}, t-t')]_{t>t'}^2\;,
\label{response1}
\end{eqnarray}
where $G_{\chi_j}^{++}({\bf q}, t-t')$ is given by
(see e.g.
\cite{br} for the explicit expressions for both the scalar and fermion field
propagators)
$G_{\chi_j}^{++}({\bf q} , t-t') = G_{\chi_j}^{>}({\bf q},t-t')
\theta(t-t') + G_{\chi_j}^{<}({\bf q},t-t') \theta(t'-t)$. 
Here $G_{\chi_j}^{>},\; G_{\chi_j}^{<}$ are

\begin{eqnarray}
&& G_{\chi_j}^> ({\bf q}, t-t') = \frac{1}{2 \omega_{{\bf q},\chi_j}(0)} \left\{
e^{-i [\omega_{{\bf q},\chi_j }(0) - i \Gamma_{\chi_j}] (t-t')} \theta(t-t') 
\right. \nonumber \\
&& \left. ~~~~~~~~~~~~~~~~~~+
e^{-i [\omega_{{\bf q},\chi_j}(0) + i \Gamma_{\chi_j}] (t-t')} \theta(t'-t) \right\}
\nonumber \\
&& G_{\chi_j}^< ({\bf q}, t-t') = G_{\chi_j}^> ({\bf q}, t'-t) \;,
\label{G><}
\end{eqnarray}
where $\omega_{{\bf q},\chi_j}(0)=\sqrt{{\bf q}^2 +m_{0\chi_j}^2 + {\rm
Re} \Sigma_{\chi_j} (q) +g_j^2 \varphi_0^2}$, with $\Sigma_{\chi_j}(q)$
the $\chi_j$ field self-energy (recall that the field decay width
$\Gamma_{\chi_j}$ is related to the imaginary part of the self-energy as
$\Gamma_{\chi_j}(q) = - {\rm Im} \Sigma_{\chi_j}(q)/(2 \omega_{{\bf
q},\chi_j})$).
Thus using Eq. (\ref{G><}) in Eq. (\ref{response1}), 
the explicit expression for
Eq. (\ref{response}) becomes

\begin{eqnarray}
\lefteqn{\!\!\!\!\!\! \langle\chi_j^2 \rangle \simeq 
\int \frac{d^3 q}{(2 \pi)^3} \frac{1}{ 2 
\omega_{{\bf q},\chi_j}(0)}   } \nonumber \\
&&  - \: g_j^2 \int_{-\infty}^t d t' \left[\varphi^2 (t') - \varphi_0^2\right] 
\int \frac{d^3 q}{(2 \pi)^3}   \nonumber \\
&& \times 2\;
\frac{\exp(-2 \Gamma_{\chi_j} |t-t'|)}{\left[2 \omega_{{\bf q},\chi_j}(0)
\right]^2} 
\sin[2\omega_{{\bf q},\chi_j}(0) |t-t'|]\;.
\label{response2}
\end{eqnarray}
{}For the second term on the RHS of Eq. (\ref{response2}),
after integrating by parts with respect to $t'$, it becomes

\begin{eqnarray}
\lefteqn{- \: g_j^2 \int_{-\infty}^t d t' \left[\varphi^2 (t') - \varphi_0^2\right] 
\int \frac{d^3 q}{(2 \pi)^3}  } \nonumber \\ 
&&\times 2 \;
\frac{\exp(-2 \Gamma_{\chi_j} |t-t'|)}{\left[2 \omega_{{\bf q},\chi_j}(0)
\right]^2} 
\sin[2\omega_{{\bf q},\chi_j}(0) |t-t'|] = \nonumber \\
&&  -\: g_j^2  \left[\varphi^2 (t) - \varphi_0^2\right] 
\int \frac{d^3 q}{(2 \pi)^3} 
\frac{1}{4 \omega_{{\bf q},\chi_j}(0) \left[\omega_{{\bf q},\chi_j}^{2}(0) +
\Gamma_{\chi_j}^2 \right]}  \nonumber \\
&& + \: g_j^2 \int_{-\infty}^t d t' \varphi(t') \dot{\varphi}(t')
\int \frac{d^3q}{(2 \pi)^3} \exp(-2 \Gamma_{\chi_j} |t-t'|)
\nonumber \\
&& \times
\frac{\left\{
\omega_{{\bf q},\chi_j}\!(0) \! \cos[2 \omega_{{\bf q},\chi_j}\!(0) |t 
\!- \!t'|] \! + \!
\Gamma_{\chi_j} \! \! \sin[2 \omega_{{\bf q},\chi_j}\!(0) |t \!- \!t'|]
\right\}}{2 \omega_{{\bf q},\chi_j}^2(0) \;
[\Gamma_{\chi_j}^2 +\omega_{{\bf q},\chi_j}^2(0)]}  .\!\!\!
\label{response3}
\end{eqnarray}

The first (local) terms on the RHS of both
Eqs. (\ref{response2}) and (\ref{response3}),
when pertubatively expanded in the coupling constant
lead to quantum corrections from the $\chi_j$-fields
to $m_{0\phi}^2$ and $\lambda$, to order $g_j^2$ and $g_j^4$,
respectively.
These corrections
are divergent but are renormalized by the usual
procedure of adding mass and coupling constant counter-terms.
The second term
on the RHS of Eq. (\ref{response3}) is responsible for
dissipation. In this study, we are interested in the regime
where $\varphi(t)$ changes slowly relative to the relaxation time,
in this case set by $\Gamma_{\chi_j}$, which means
the adiabatic approximation is valid.
Under this approximation, similar to the treatment in \cite{br},
a Markovian, or equivalently time local, treatment can be
used, which amounts to a derivative expansion of the field
$\varphi(t)$ and in which the leading ${\dot \varphi}$
term  only is retained.  After implementing this approximation
and substituting Eq. (\ref{response3}) back
into Eq. (\ref{response2}), we obtain
\begin{eqnarray}
\langle \chi_j^2 \rangle &\simeq&
\int \frac{d^3 q}{(2 \pi)3 2 \omega_{{\bf q},\chi_j}(t)}
\left\{1+  \frac{g_{j}^2 \varphi \dot{\varphi} \Gamma_{\chi_j}}
{\left[\omega_{{\bf q},\chi_j}^{2}(t) +
\Gamma_{\chi_j}^2 \right]^2} \right\}\;.
\label{averchi2}
\end{eqnarray}
In the above, note we have conveniently reintroduced
the time dependence back into the field frequencies and
when they are perturbatively expanded
to order $g_j^4$, the above mentioned mass and coupling constant corrections
are correctly reproduced.

An analogous expression to Eq. (\ref{averchi2}) also follows for
$\langle \phi^2 \rangle$. Note however that for an initial (at $t=t_0$)
zero temperature bath and for fields $\Phi$ and $\chi_j$ satisfying the
mass constraint $m_{\chi_j} > 2 m_{\psi_k} > m_{\phi}$, there only will
be decay channels for $\chi_j$ into $\psi_k$ particles. As a result, it
implies $\Gamma_\phi (q)=0$ and $\Gamma_{\psi_k}(q)=0$, while we have that

\begin{equation}
\Gamma_{\chi_j}(q)= \sum_{k=1}^{N_\psi} h_{kj}^2
\frac{m_{\chi_j}^2}{8\pi  \omega_{{\bf q},\chi_j}}
\left(1-\frac{4 m_{\psi_k}^2}{ m_{\chi_j}^2}\right)^{3/2}\;.
\label{rate}
\end{equation}
As such, in the adiabatic regime, dissipation will only involve the
decay of $\chi_j$ particles. The other two averages in the EOM, $\langle
\phi^3 \rangle$ and $\langle \phi \chi_j^2 \rangle$ can also be worked
out in the linear response approach, and
their leading contributions are at two-loop order \cite{br}. Here, we will
not consider them but restrict our calculation to leading
one-loop order for simplicity.  In this
case, the only contribution to dissipation is
Eq. (\ref{averchi2}), and this effect
already will be adequate to demonstrate
considerable radiation production from our model Lagrangian. 
Substituting  Eq. (\ref{averchi2}) back into the
effective EOM, Eq. (\ref{eqphi}), the second term on the RHS of Eq.
(\ref{averchi2}) leads to a dissipative term in the EOM and the first
term leads to $\Phi$ mass and coupling constant divergent corrections, 
that can be renormalized as usual by the introductions of counterterms
in Eq. (\ref{model}). This renormalization procedure is standard and
will not be further addressed. In our final expressions, all mass
parameters, $m_{0\phi}$, $m_{0\chi_j}$,$m_{0\psi_k}$, and coupling
constants, $\lambda$,$g_j$,$h_{kj}$ are then taken as the renormalized
ones. The renormalized effective EOM for $\varphi(t)$ that finally
emerges can be written as

\begin{eqnarray}
\ddot{\varphi} + \frac{\partial V_{\rm eff}(\varphi)}{\partial \varphi}
+ \eta(\varphi) \dot{\varphi}=0 \;.
\label{eqphi2}
\end{eqnarray}
In the above equation, we have included in $V_{\rm eff}$ the 
quantum renormalization corrections to the mass
and coupling constant for the $\Phi$ field, which are exactly
the same as found
in the calculation of a constant background 
$\varphi$-field effective potential.
The dissipation coefficient $\eta(\varphi)$ in Eq. 
(\ref{eqphi2}) comes from 
performing the momentum integral in Eq. (\ref{averchi2})
and using (\ref{rate}) to give
\begin{eqnarray}
\eta (\varphi)  =  
\varphi^2(t) \sum_{j=1}^{N_\chi} \frac{ g_j^4  
\alpha_{\chi,\psi}^2 
\left(m_{\chi_j}^4 + \alpha_{\chi,\psi}^4\right)^{-1/2}}  
{32 \pi \left(2 \sqrt{m_{\chi_j}^4 + \alpha_{\chi,\psi}^4} + 
2m_{\chi_j}^2 \right)^{1/2}}\;,
\label{eta}
\end{eqnarray}
where $ \alpha_{\chi,\psi}^2 = \sum_{k=1}^{N_\psi} h_{kj,\chi}^2
m_{\chi_j}^2 \left(1-4 m_{\psi_k}^2/m_{\chi_j}^2 \right)^{3/2}/(8 \pi) $
and $m_{\chi_j}$ in Eq. (\ref{eta}) denote the field dependent masses,
$m_{\chi_j}^2 \equiv m_{\chi_j}^2 (\varphi) =m_{0\chi_j}^2 + g_j^2
\varphi^2(t)$. 
The dissipative mechanism Eq. (\ref{eta})
overcomes an underlying impediment to
realizing robust warm inflation in the finite temperature calculations
\cite{bgr1,yl}, where all mass scales were constrained
by the temperature.
In sharp contrast, a key
feature about the dissipative mechanism of this paper is
that irrespective of the magnitude of $\varphi$ and $m_{\chi_j}$,
dissipation occurs unchecked by these severely limiting constraints.

{}For the dissipative
mechanism derived in this letter to be applicable to warm inflation,
there must be some control in determining the quantum corrections in
$V_{\rm eff}$ in Eq. (\ref{eqphi2}). This is required mainly since, similar
to supercooled inflation, in the warm inflation case also, treatment of
density perturbations requires an ultraflat potential
\cite{wi,phenomen,bf2ab2}. 
However, there are one-loop quantum corrections to 
the ${\rm T} = 0$ effective potential
arising in the Lagrangian Eq. (\ref{model}) from the self-interaction
of the $\phi$-field and from its interactions with the
$\chi$-fields, which give \cite{dj}
\begin{equation}
V_1(\varphi) =\frac{1}{2} \int \frac{d^3k}{(2\pi)^3}
\left( E_{m_{\phi}} + \sum_{i=1}^{N_{\chi}}  E_{m_{\chi_i}} \right) ,
\end{equation}
where
$E_{m_{\phi}}  = \sqrt{{\bf k}^2 + m_{0\phi}^2 + \lambda \varphi^2/2}$
and
$E_{m_{\chi_i}}  =  \sqrt{{\bf k}^2 + m_{0\chi_i}^2 + g_i^2 \varphi^2}$.
To obtain the desired ultraflat potential, it requires $\lambda$ to
be tiny with $m^2_{0\phi} \stackrel{<}{\sim} \lambda \varphi^2/2$.
In this regime, the contribution from the $E_{m_{\phi}}$ term
above is negligible.  However, since in general we want
$g_i^4 \gg \lambda$, the one-loop contributions from
the $E_{m_{\chi_i}}$ terms lead to corrections $ \sim g_i^4 \varphi^4$
in $V_{\rm eff}$ and thus would ruin the flatness of the potential.
Operationally these one-loop contributions can be controlled
by adding to the Lagrangian Eq. (\ref{model}) fermionic 
``partners'' $\psi^{\chi}$ to the $\chi$-fields, with one
$\psi^{\chi}$-field for every four $\chi$-fields and
coupling only to the $\Phi$-field as
$\sum_{i=1}^{N_{\chi}/4} g^{\chi}_i{\bar \psi}^{\chi}_i \psi^{\chi}_i \Phi$.
The one-loop quantum corrections to the effective
potential from these terms will yield \cite{dj}

\begin{equation}
V_1(\varphi) =-2 \int \frac{d^3k}{(2\pi)^3}
\sum_{i=1}^{N_{\chi}/4}  E_{m_{\psi^{\chi}_i}},
\end{equation}
where
$E_{m_{\psi^{\chi}_i}}  =  \sqrt{k^2 + (m_{0\psi^{\chi}_i} + g^{\chi}_i \varphi)^2}$.
In particular, this fermionic contribution has the familiar
opposite sign to the bosonic contribution.
Thus with appropriately tuned parameters
$g_i$, $g^{\chi}_i$ and with zero explicit masses
$m_{0\psi_{\chi_i}}=m_{0\chi_i}=0$, the one-loop quantum corrections to
$V_{\rm eff}$ cancel to all orders in $g_i$, $g^{\chi}_i$.
This modification
simply is mimicking supersymmetry. {}For realistic model building, the
mechanism derived in this letter must be examined in actual SUSY models,
where the choice $g_i^4 \gg \lambda$ of coupling parameters 
could be obtained naturally, 
but that will not be pursued here.

\section{Alternative derivation of dissipation - operator formalism}
\label{sect3}

{}For completeness, here an alternative derivation of dissipation
is presented using the canonical approach and following the formalism
developed in \cite{ms1,ring1}.  In this approach,
the fields $\phi$, $\chi$ and $\psi$ are expressed in terms of their mode
decompositions and dynamics is determined with respect to the
mode operators.
Thus, for example for the $\chi_j({\bf x},t)$ field this means

\begin{equation}
\chi_j({\bf x},t) = \int \! \frac{d^3 q}{(2 \pi)^\frac{3}{2} 
[2 \omega_{{\bf q},\chi_j} (t)]^\frac{1}{2}}
\left[ a_{{\bf q},\chi_j} (t) e^{-i{\bf q}.{\bf x}} + 
a^\dagger_{{\bf q},\chi_j} (t)
e^{i{\bf q}.{\bf x}}
\right]\;.
\label{chia}
\end{equation}
Since there is a time dependent background field $\varphi(t)$,
this induces time dependence in the frequencies and so in the
creation/annihilation operators of the $\phi$ and
$\chi_j$ fields.
In the analysis that follows, we will focus on the
$\chi_j$ fields, with similar considerations carrying over for the
$\phi$ field.

The time dependent
$\chi_j$ - frequency in Eq. (\ref{chia}) is given by
$ \omega_{{\bf q},\chi_j} (t) = [{\bf q}^2 + m_{0\chi_j}^2 +
g_j^2 \varphi^2 (t)]^{1/2}$. 
{}From Eq. (\ref{chia}) it follows that

\begin{eqnarray}
\langle \chi_j^2 \rangle =
\int \frac{d^3 q}{(2 \pi)^3 2 \omega_{{\bf q},\chi_j}(t)}
\left[2 x_{{\bf q},\chi_j}(t) + 2  {\rm Re} [y_{{\bf q},\chi_j}(t)] +1\right] ,
\label{averchi}
\end{eqnarray}
where
$x_{{\bf q},\chi_j}(t)= \langle a_{{\bf q},\chi_j}^{\dagger}(t)
a_{{\bf q},\chi_j}(t) \rangle$
is the particle
number density and
$y_{{\bf q},\chi_j}(t)= \langle a_{{\bf q},\chi_j}(t) a_{-{\bf q},\chi_j}(t)
\rangle$ is the
off-diagonal correlation.

{}From the field equation for $\chi_j$ and Eq. (\ref{chia}) we can
deduce the equations satisfied by $ x_{{\bf q},\chi_j}$ and
$y_{{\bf q},\chi_j}$. Taking also into account the possibility that the field
$\chi_j$ can decay into lighter fields with a decay rate
$\Gamma_{\chi_j}(q)$ as already given in Eq. (\ref{rate}), 
$ x_{{\bf q},\chi_j}$ and $y_{{\bf q},\chi_j}$ can be shown to
satisfy the coupled differential equations \cite{ms1,ring1}

\begin{eqnarray}
{\dot x}_{{\bf q},\chi_j} &=& \frac{\dot{\omega}_{{\bf q},\chi_j}}{\omega_{{\bf q},\chi_j}}
{\rm Re} \, y_{{\bf q},\chi_j} \;,
\nonumber \\
\dot{y}_{{\bf q},\chi_j} &=& \frac{\dot{\omega}_{{\bf q},\chi_j}}{\omega_{{\bf q},\chi_j}
- i \Gamma_{\chi_j}(q)}
\left[ x_{{\bf q},\chi_j} +\frac{1}{2} \right] \nonumber \\
&-& 2 i \left[
\omega_{{\bf q},\chi_j}- i \Gamma_{\chi_j}(q) \right]
y_{{\bf q},\chi_j} \;.
\label{diff}
\end{eqnarray}
A solution for Eq. (\ref{diff}) can be found in the quasi-adiabatic
regime as follows. 
Let us consider the case of a slowly changing configuration
$\varphi(t)$. We can therefore suppose
that the number of produced particles at time $t$ is
$x_{{\bf q},\chi_j}(t)  \ll 1$. Consequently we also have
that
$\omega_{{\bf q},\chi_j}$ and its time derivative
slowly change. We then find for $y_{{\bf q},\chi_j}$ in Eq. (\ref{diff}) the result

\begin{eqnarray}
y_{{\bf q},\chi_j}(t)&=&- i\,\frac{\dot{\omega}_{{\bf q},\chi_j}
\left\{ 1-\exp\left[-2
i\left(\omega_{{\bf q},\chi_j}-i\Gamma_{\chi_j} \right) t \right]  \right\} }{4\left[
\omega_{{\bf q},\chi_j}-i\Gamma_{\chi_j}(q)\right]^2}\;,
\label{result1}
\end{eqnarray}
which in the limit $t\gg\Gamma_{\chi_j}^{-1}$ yield

\begin{eqnarray}
{\mathrm{Re}}\,y_{{\bf q},\chi_j}(t)&=&
\frac{g_{j}^2}{2} \varphi \dot{\varphi}
\frac{\Gamma_{\chi_j}}
{\left(\omega_{{\bf q},\chi_j}^{2}+
\Gamma_{\chi_j}^2 \right)^2}\;.
\label{result2}
\end{eqnarray}

Using Eq. (\ref{result2}) in Eq. (\ref{averchi}), once again we get Eq.
(\ref{averchi2}), from which the effective EOM Eq. (\ref{eqphi2})
follows. A shortcoming of this approach is that interactions are
added to the set of Eqs. (\ref{diff}) in a somewhat ad-hoc way.
This point was discussed recently
in \cite{ian2}, where the complete kinetic equations where derived
for the single field self-interacting $\phi^4$ model.
Nevertheless, the final answer from the approach of this section agrees
with that from the Lagrangian based approach of the previous
section, where interactions can be added consistently through the
appropriate set of Schwinger-Dyson equations for the propagators \cite{br}.
Thus it suggests the results by this canonical approach are acceptable,
but missing gaps in the formalism of \cite{ms1} must still
be resolved.  {}For our purposes, due to the importance of the dissipative 
mechanism studied in this letter, we felt it was important to point
out the agreement between independently developed formalisms,
even if there remain shortcomings in one of them.
The practical significance of the results in this letter
provide motivation to address these difficult problems
in the course of future work.

\section{Physical interpretation and an explicit application}
\label{sect4}

We now turn to an application of the equations derived above, using an
explicit set of model parameter values, which are consistent with simple
inflationary models. But before that, let us address briefly the physical
interpretation of dissipation in Eq. (\ref{eqphi2}).

We note that the evolving background field $\varphi(t)$ changes the
masses of the $\chi_j$ bosons. As a consequence, the positive and
negative frequency components of the $\chi_j$-fields mix. This in turn
results in the coherent production of $\chi_j$ particles which then
decohere through decay into lighter $\psi_k$-fermions. This picture can
be confirmed by checking energy balance. This is done by examining the
time evolution of the $\chi_j$-particle number density. {}For this,
their number density is expressed in terms of time dependent creation
and annihilation operators as ${\cal N} \equiv \sum_j \langle
a_{\chi_j}^{\dagger}(t) a_{\chi_j}(t) \rangle$. By relating the time
dependent operators $a_{\chi_j}^{\dagger}(t)$ and $a_{\chi_j}(t)$ to the
initial, time independent, creation and annihilation operators through a
Bogoliubov transformation, the total particle production rate then can
be computed in general.  Thus, the time evolution of the
total production rate is
\begin{equation}
\dot{\cal N} =\sum_{j=1}^{N_\chi} \int \frac{d^3 k}{(2 \pi)^3} 
{\dot x}_{{\bf q},\chi_j}\;,
\end{equation}
which using Eqs. (\ref{diff}) and (\ref{result2}),
leads to
\begin{eqnarray}
\dot{{\cal N}}& = &\dot{\varphi}^2 \sum_{j=1}^{N_\chi} 
\int \frac{d^3 k}{(2 \pi)^3}
\frac{g_{j}^4}{2 \omega_{{\bf q},\chi_j}}
\frac{\Gamma_{\chi_j}}
{\left(\omega_{{\bf q},\chi_j}^{2}+
\Gamma_{\chi_j}^2 \right)^2} \;.
\label{ndot}
\end{eqnarray}
It can now be checked from Eqs. (\ref{eqphi}), (\ref{averchi2})
and (\ref{eqphi2}), that the above result, Eq. (\ref{ndot}) is precisely 
equal to the vacuum energy loss rate, $\eta {\dot \varphi}^2$, 
as obtained from the effective EOM,
Eq. (\ref{eqphi2}).

Let us now examine the application of the results in this letter
to warm inflation and also understand their significance.
The scope of the present calculation is limited since
dissipation at zero temperature necessarily implies
a nonequilibrium state, which is evolving to some
statistical state containing particles.  Thus the estimates
made below only give some idea of the magnitude of particle
production.  However, provided the magnitude is significant,
as will be shown, it reveals that on scales relevant to
inflation, quantum field theory with generic interactions
has robust tendency to dissipate.  
{}For our estimates, we have set same all $\Phi - \chi$ couplings
$g_{\chi_j} = g$  as well all
$\chi-\psi$ couplings, $h_{kj} = h$.

We are interested in overdamped motion for the inflaton $\varphi(t)$,
which requires i). $m^2_{\phi} \equiv m^2_{\phi}(\varphi) = m_{0\phi}^2 +
\lambda \varphi^2/2 \; < \; \eta^2(\varphi)/4$. The adiabatic
approximation Eq. (\ref{adapp}) requires
ii). $m^2_{\phi}(\varphi)/\eta(\varphi) < \Gamma_{\chi}$. Although our
derivation was for Minkowski spacetime, provided the time scale of
microscopic dynamics is faster than the Hubble time scale, then within
sub-Hubble length scales, this Minkowski spacetime calculation should be
valid. {}For this to hold, it requires iii). $H=
\sqrt{8\pi V_{\rm eff} /3m_{\rm pl}^2} \simeq \sqrt{8\pi (\lambda/4!)\:
\varphi^4/3m_{\rm pl}^2} < \Gamma_{\chi}$, where $m_{\rm pl}$ is the Planck mass. 
Also, so that the macroscopic
motion of $\varphi$ is governed by the dissipative term it requires
iv). $\eta(\varphi) > 3 H$. Thus combining all four of the above consistency conditions
leads to parametric constraints.
To obtain these, we will treat $m_{\chi} \geq \alpha_{\chi, \psi}$,
where from below Eq. (\ref{eta}) we have, by setting $m_{\chi}^2 \sim g^2 \varphi^2$, 
$\alpha_{\chi,\psi}^2 \approx g^2 h^2 N_{\psi} \varphi^2/(8 \pi)$,
which thus requires $h^2 N_{\psi}/(8 \pi) < 1$.  In this regime, we have from 
Eq. (\ref{eta}) $\eta \approx g^3 h^2 N_{\chi} N_{\psi} \varphi/(512 \pi^2)$
and from Eq. (\ref{rate}) $\Gamma_{\chi} \approx g h^2 N_{\psi}\varphi/(8\pi)$.
The parametric constraints that follow from the four conditions
given above are respectively
\begin{eqnarray} 
\label{lamcon}
{\rm i. \ } & & \lambda < \frac{g^6 h^4 N_{\chi}^2 N_{\psi}^2}{2(512 \pi^2)^2}
\nonumber \\
{\rm ii. \ } & & \lambda < \frac{g^4 h^4 N_{\chi} N_{\psi}^2}{2048 \pi^3}
\nonumber \\
{\rm iii. \ } & & \lambda < \frac{9 g^2 h^4 N_{\psi}^2}{64 \pi^3}
\frac{m_{\rm pl}^2}{\varphi^2}
\nonumber \\
{\rm iv. \ } & & \lambda < \frac{g^6 h^4 N_{\chi}^2 N_{\psi}^2}{\pi(512 \pi^2)^2}
\frac{m_{\rm pl}^2}{\varphi^2} \;.
\end{eqnarray}
To yield large dissipation, we are usually interested in the regime where
the couplings $g, h$ are big.  To remain within a well defined perturbative
region, we will then further require that $g^4 N_{\chi} \lesssim 1$ 
and $h^2 N_{\psi} \lesssim 1$ and will base our
estimates on the upper bounds here.  Also, in general
$\varphi \lesssim m_{\rm pl}$, but to obtain the tightest
constraints on $\lambda$ in (iii) and (iv), we will set this at the
equality point.  Under these conditions, we find
for the constraints (i) - (iv) in Eq. (\ref{lamcon}) respectively
$\lambda < {\rm min}( 10^{-8} g^2 N_{\chi}, 10^{-5}, 10^{-3} g^2 ,10^{-8}g^2 N_{\chi} )$.
Recalling that constraints imposed by density fluctuations
give typically $\lambda < 10^{-14}$ \cite{wi,phenomen,bf2ab2}, we see that the
above constraints introduce no stricter limitations.

As shown in Eq. (\ref{ndot}), radiation production is determined by
\begin{eqnarray}
{\dot \rho}_r(t) & = & \eta(\varphi) {\dot \varphi}^2
= - \frac{dV_{\rm eff}}{d \varphi} {\dot \varphi}
\approx V_{\rm eff}(\varphi) \frac{m^2_{\phi}(\varphi)}{\eta}.
\label{rhodot}
\end{eqnarray}
The zero temperature calculation should be valid for a time period
$\sim 1/\Gamma_{\chi}$, in which time the magnitude of radiation
produced is

\begin{equation}
\label{rhogam}
\rho_r(1/\Gamma_{\chi}) \approx V_{\rm eff}(\varphi)
m^2_{\phi}(\varphi)/(\eta \Gamma_{\chi}) < V_{\rm eff}(\varphi).
\label{rad}
\end{equation}
Based on Eqs. (\ref{rate}) and (\ref{eta}) and the above constraints on
$\lambda$, there is considerable freedom in choosing the ratio
${\cal R} \equiv m^2_{\phi}/(\eta \Gamma_{\chi})$
appearing in Eq. (\ref{rad}). Considering then
an ultraflat potential, as necessary for observationally consistent
density perturbations, which requires typical values of $\lambda \lesssim
10^{-14}$, this implies ${\cal R} \lesssim
10^{-10}/(g^4h^4 N_{\psi}^2 N_{\chi})$. {}For unexceptional values of the
perturbative coupling parameters, say $g \sim h \sim 0.1$, and small
number of $\chi$ and $\psi$ fields, $N_{\chi}, N_{\psi} \sim 1 - 10$,
this leads to ${\cal R} \sim 10^{-(2-5)}$. Also note these parameters choices
are consistent with the conditions on $\lambda$ given above Eq.
(\ref{rhodot}). Thus for a typical scale for inflation, where the
potential energy is at the GUT scale, 
$V_{\rm eff}(\varphi)^{1/4} \sim 10^{15-16}
{\rm GeV}$, it implies a generated radiation component which, if
expressed in terms of temperature, is at the scale $T \sim 10^{13-16}
{\rm GeV}$, and this is non-negligible. This is a significant result not
only because the magnitude of produced radiation is large, but also
because it emerges from a very generic interaction, scalar $\rightarrow$
heavy scalar $\rightarrow$ light fermions, which is very common in many
particle physics models. Moreover, we expect similar robust radiation
production for decay of the heavy scalars into gauge bosons. {}Finally,
although we did this zero temperature calculation first simply due to
its tractability, an interesting fact emerges for inflationary
cosmology, that even if the initial state of the universe before
inflation is at zero temperature, the dynamics itself could bootstrap
the universe to a higher temperature during inflation.

\section{Extension to expanding space-time}
\label{sect5}

The extension of this calculation is formally straightforward to 
Friedmann-Robertson-Walker (FRW)
spacetime, $ds^2 = dt^2 - a^2(t) d{\bf x}^2$, where $a(t)$ is the cosmic
scale factor and $t$ is cosmic time. 
In this case, the extension of Eq. (\ref{model}), for the Lagrangian 
density of the matter fields coupled to the gravitational field
tensor $g_{\mu \nu}$, is given by  
 
\begin{eqnarray} 
\lefteqn{ {\cal L} = 
\sqrt{-g} \left\{ \frac{1}{2} g^{\mu \nu} 
\partial_\mu \Phi \partial_\nu \Phi - \frac{m_{0\phi}^2}{2}\Phi^2 - 
\frac{\lambda}{4 !} \Phi^4  -\frac{\xi}{2} R \Phi^2 
\right. } \nonumber \\
&& \!+ \!\!\left. \sum_{j=1}^{N_{\chi}} \left[ 
\frac{g^{\mu \nu}  }{2} \partial_\mu \chi_{j} \partial_\nu \chi_{j}\! -\! 
\frac{m_{0\chi_j}^2}{2}\chi_j^2 
\!-\! \frac{f_{j}}{4!} \chi_{j}^4 \!-\! \frac{g_{j}^2}{2} 
\Phi^2 \chi_{j}^2  \!-\! \frac{\xi}{2} R \chi_j^2
\right] \right.
\nonumber \\ 
&& \!\!+\! \left. \sum_{k=1}^{N_{\psi}} \! \left[ 
\!i \bar{\psi}_{k} \gamma^\mu \! \left(\! \partial_\mu \!+\! \omega_\mu \!
\right)\psi_k 
\!-\! \bar{\psi}_k \!\!\left(\! m_{0\psi_k} \!\! +\! \! 
\sum_{j=1}^{N_\chi} h_{kj} \chi_j \! \right) \! \psi_k \!\right]\! \right\}, 
\label{Nfieldsfrw} 
\end{eqnarray} 
 
\noindent  
where $R$ is the curvature scalar and $\xi$ is the dimensionless parameter
describing the coupling of the matter fields to the gravitational background. 
In the last terms involving
the fermion fields, the $\gamma^\mu$ matrices are related to the vierbein
$e_\mu^a$ (where $g_{\mu\nu}=e_\mu^a e_\nu^b \eta_{ab}$, with $\eta_{ab}$
the usual Minkowskii metric tensor) by $\gamma^\mu (x) =
\gamma^a e_a^\mu (x)$ \cite{mallik}, where $\gamma^a$ are the usual Dirac matrices
and $\omega_\mu = -(i/4) \sigma^{ab} e_a^\nu \nabla_\mu e_{b\nu}$, with
$\sigma^{ab} =i/2 [\gamma^a,\gamma^b]$.

It is easy to show that the
Lagrangian Eq. (\ref{Nfieldsfrw}) in conformal time, $t_c$, where $dt= a
dt_c$, remains unchanged from Eq.(\ref{model})
except that all masses obtain time dependence
related to $a(t_c)$ (see for example \cite{ring1} for a similar
problem). In particular, for the bosonic fields we have that
$m_{\chi_j}^2(t_c) = m_{0\chi_j}^2 a^2(t_c) - d^2a/2adt_c^2+ \xi a^2
R/2$ and similar for the $\phi$ field, and for the fermionic fields
$m_{\psi_k}(t_c)= m_{0\psi_k}a(t_c)$. These time dependent
parameters can be treated within the linear response formalism used in
this letter. Moreover, since the time dependence is associated with
$a(t_c)$, it is easy to show that provided $H < \Gamma_{\chi}$, the time
dependence of the mass terms is slow relative to microscopic dynamics
and thus an appropriate adiabatic approximation should be applicable. 

The observations made above are adequate to establish that, for the
mechanism of central interest in this letter, the robust dissipative
properties found above for Minkowski spacetime also will hold for
expanding spacetime. However, the exact form of the effective
$\varphi$-EOM is a more involved matter. The problem is there are three
relevant time scales $H$, $\Gamma_{\chi_j}$ and ${\dot
\varphi}/{\varphi}$, where for the slow-roll motion of interest, we seek
solutions with ${\dot \varphi}/\varphi < H$. Moreover, ultimately we
require the evolution equation in cosmic time, and the relation between
that and conformal time is in general very nonlinear. {}For example, for
the case of prime interest, de Sitter space, $t \propto \ln (1- bt_c)$.
Thus power law ambiguities can have nontrivial relevance in relating
between conformal and cosmic time, and such ambiguities are prevalent in
adiabatic approximations and derivative expansions. This is a serious
matter and to learn more about this mechanism in expanding spacetime
beyond what already has been understood from the above Minkowski
spacetime calculation requires application of more complete
nonequilibrium methods, such as \cite{idl}. We will consider the details
of this derivation in the FRW spacetime in a future work.
  
\section{Conclusions}
\label{concl}

The relevance of the analysis in this letter extends beyond warm
inflation, since the interactions studied here are exactly the same as
found in supercooled inflation models. In fact, in the context
of the model studied here, with couplings around the ones studied in the
example of Sec. \ref{sect4}, reheating becomes irrelevant,
since our analysis showed the model
is inconsistent with supercooling in the first stage, and the
entire dynamics is warm throughout. Thus, as 
originally suggested \cite{wi,bf2ab2}, warm inflation dynamics is
inherently intertwined with the general problem of inflationary
dynamics.

Since the first principle results in this paper
give support to the warm inflation picture, it is worth
recalling here other features that also have made this picture
compelling.  {}First, warm inflation
overcomes a conceptual barrier that the supercooled picture
has never shaken away, which is that in warm inflation there is no
quantum-classical transition problem, since the macroscopic dynamics of
the background field and fluctuations \cite{bf2ab2} are classical from
the onset.  Second, in warm inflation models, in
regimes relevant to observation,
the mass of the inflaton field
is typically much larger than the Hubble scale, thus
these models do not suffer from what is sometimes called
the ``eta problem''.
Finally, accounting for dissipative effects may be important
in alleviating the initial condition problem of inflation \cite{bg,ROR}.

The emerging picture is that warm inflation remains a hopeful direction
toward a complete and consistent dynamical description of the
early universe.  However, considerable work remains in understanding
the quantum field theory of this picture.  Two areas were already
identified in the paper.  One is resolving the gaps in
the canonical dissipative formalism of \cite{ms1}, thus
permitting this approach to be a viable cross-check to the Lagrangian
approach.  The other area is a full investigation of the dissipative
formalism in expanding spacetime.  Beyond this, the more difficult
problem is extending the adiabatic contraints in the present
formalisms to treat nonequilibrium conditions.  Steps along this
direction already have begun, using operator methods \cite{HR}
and the even more ambitious attempt in \cite{ian2}
to derive the Boltzmann-like kinetic equation for
interacting fields.

\acknowledgements

It is a pleasure to thank Ian Lawrie and Marc Sher for helpful discussions. 
A.B. is funded by the United Kingdom
Particle Physics and Astronomy Research Council (PPARC) and R.O.R.
is partially supported by Conselho Nacional 
de Desenvolvimento Cient\'{\i}fico e
Tecnol\'ogico (CNPq-Brazil).


\begin{thebibliography}{99}

\bibitem{olive} K. A. Olive, Phys. Rep. {\bf 190} (1990) 307.

\bibitem{wi} A. Berera,  Phys. Rev. Lett. {\bf 75}
(1995) 3218;
Phys. Rev. D{\bf 54} (1996) 2519;
Phys.\ Rev.\  D{\bf 55} (1997) 3346.

\bibitem{phenomen}
H. P. de Oliveira and  R. O. Ramos, 
Phys. Rev. D{\bf 57} (1998) 741;
W. Lee and  L.-Z. Fang, Phys. Rev. D{\bf 59} (1999) 083503;
J. M. F. Maia and J. A. S. Lima, Phys. Rev. D{\bf 60}
(1999) 101301;  M. Bellini, Class. Quant. Grav. {\bf 16} (1999) 2393;
A. N. Taylor and  A. Berera, Phys. Rev. D{\bf 62}
(2000) 083517;  
H. P. De Oliveira and S. E. Joras, Phys. Rev. D{\bf 64} (2001) 063515;
L. P. Chimento, A. S. Jakubi, D. Pavón
and N. A. Zuccala, Phys. Rev. D{\bf 65} (2002) 083510.

\bibitem{bgr1} A. Berera, M. Gleiser and R. O. Ramos, Phys. Rev. D{\bf 58}
(1998) 123508.

\bibitem{bgr2} A. Berera, M. Gleiser and R. O. Ramos,
Phys. Rev. Lett. {\bf 83} (1999) 264.

\bibitem{yl} J. Yokoyama and  A. Linde,
Phys. Rev. D{\bf 60} (1999) 083509.

\bibitem{br}A. Berera and R. O. Ramos, Phys.
Rev. D{\bf 63} (2001) 103509.

\bibitem{ian2} I. D. Lawrie, Phys. Rev. D{\bf 66} (2002) 041702.

\bibitem{HR}  G. Flores-Hidalgo and R. O. Ramos,
hep-th/0206022 (in press Physica A).

\bibitem{gr} M. Gleiser and R. O. Ramos, Phys. Rev. D{\bf 50} (1994) 2441.

\bibitem{morikawa}
M. Morikawa and M. Sasaki, Phys. Lett. B{\bf 165} (1985) 59;
M. Morikawa, Phys. Rev. D{\bf 33} (1986) 3607.

\bibitem{ms1} M. Morikawa and S. Sasaki, 
Prog. Theor. Phys. {\bf 72} (1984) 782.

\bibitem{hs} A. Hosoya and M. Sakagami, Phys. Rev. D{\bf 29}
(1984) 2228 (1984).

\bibitem{ring1} A. Ringwald,  Ann. Phys. (N.Y.) {\bf 177} (1987) 129.

\bibitem{bf2ab2} A. Berera and L. Z. Fang, Phys. Rev. Lett. {\bf 74}
(1995) 1912;
A. Berera, Nucl. Phys. B{\bf 585}
(2000) 666.

\bibitem{dj} L. Dolan and R. Jackiw, Phys. Rev. D{\bf 9} (1974) 3320.

\bibitem{mallik} N. Banerjee and S. Mallik, Phys. Rev. D{\bf 45} (1992) 701.

\bibitem{idl} I. D. Lawrie, Phys. Rev. D{\bf 40} (1989) 3330.

\bibitem{bg} A. Berera and C. Gordon, Phys. Rev. D{\bf 63}
(2001) 063505.

\bibitem{ROR} R. O. Ramos, Phys. Rev. D{\bf 64} (2001) 123510.


\end{thebibliography}
\end{document}